\documentclass[
prl, 
7pt,
twocolumn,
en,
longbibliography
]{revtex4-1}

\usepackage{graphicx}
\usepackage{textcomp} 
\usepackage{gensymb} 
\usepackage{dcolumn}
\usepackage{bm}
\usepackage{amsmath, amssymb}
\usepackage{dsfont}
\usepackage{float}
\usepackage{afterpage}
\usepackage{xcolor}
\usepackage[normalem]{ulem}
\usepackage[english]{babel}

\newcommand{\ket}[1]{\vert #1 \rangle}
\newcommand{\bra}[1]{\langle #1 \vert}

\newcommand{\nbar}{\bar{n}}

\begin{document}


\title{{A Physical Quantum Agent}}

\author{M. J. Kewming}
\email{m.kewming@uq.edu.au}
\author{S. Shrapnel}%
 \author{G. J. Milburn}
\email{milburn@physics.uq.edu.au}
\affiliation{Centre for Engineered Quantum Systems, School of Mathematics and Physics, University of Queensland, Queensland, 4072 Australia}

\date{\today}
\begin{abstract}
The concept of an embodied intelligent agent is a key concept in modern artificial intelligence and robotics. 
Physically, an agent is an open system embedded in an environment that it interacts with through sensors and actuators. 
It contains a learning algorithm that correlates the sensor and actuator results by learning features about its environment. 
{In this article we present a simple optical agent that uses light to probe and learn components of its environment. In our scenario, the quantum agent outperforms a classical agent: The quantum agent probes the world using single photon pulses, where its classical counterpart uses a weak coherent state with an average photon number equal to one. We analyze the thermodynamic behavior of both agents, showing that improving the agent's estimate of the world corresponds to an increase in average work done on the sensor by the actuator pulse. Thus, our model provides a useful toy model for studying the interface between machine learning, optics, and statistical thermodynamics. } 
\end{abstract}

\maketitle

\section{Introduction}

Machine learning (ML), artificial intelligence (AI), and quantum physics are current hotbeds of academic research.
It is unquestionable that the physical sciences have benefited tremendously by incorporating the tools of ML \cite{carleo_machine_2019}.
Recent research has proposed enhancing ML tools and techniques using quantum physics \cite{schuld_introduction_2015, biamonte_quantum_2017, schuld_quantum_2019}.
However, more topical research suggests that ML on an industrial scale can have very tangible thermodynamic costs \cite{strubell_energy_2019}.
This suggests it is imperative that we develop a deeper understanding of the thermodynamic cost of learning. 
Recent results using stochastic thermodynamics have obtained fundamental bounds on the efficiency of learning algorithms \cite{goldt_stochastic_2017, goldt_thermodynamic_2017}. Others have also shown that learning maximizes the work done by a Maxwell's demon \cite{boyd_thermodynamic_2020}. 
This suggests that learning, like other physical process may take place out of thermal equilibrium \cite{england_dissipative_2015}. 
For a physical agent to learn it must interact with the world via physical sensors and actuators. 
A sensor is a physical device which the agent can use to read in information about the world. 
An actuator is a physical device which the agent can use to write information out into the world.
The agent also contains a learning algorithm that correlates the sensor and actuator results by learning features about its environment.
{Given the known advantages of quantum metrology \cite{giovannetti_advances_2011, degen_quantum_2017}, we consider an advantage that a simple AI endowed with quantum mechanical hardware---otherwise referred to as a quantum agent---may yield.}

\begin{figure*}
    \centering
    \includegraphics[width=2\columnwidth]{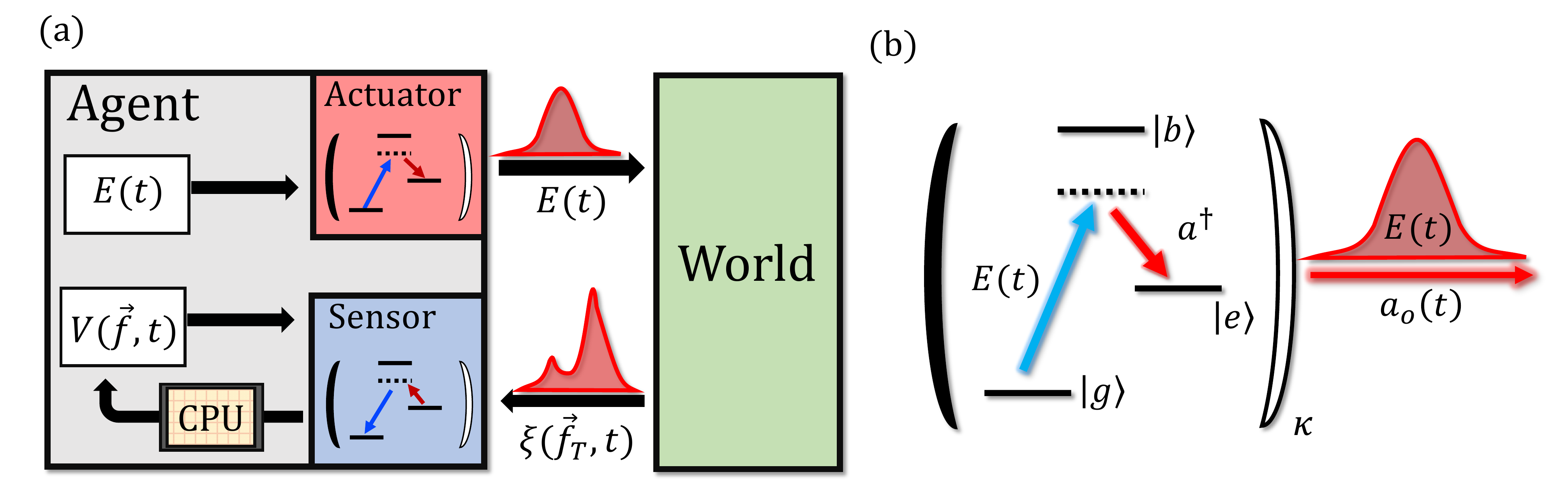}
    \caption{(a) Depiction of a learning agent interacting with the world. The agent houses both an actuator and sensor, which it uses to probe and measure the world. The actuator is maintained at a positive temperature equilibrium state (red), while the sensor is maintained in a negative temperature equilibrium state (blue). The agent emits photons with a temporal profile $E(t)$, which is perturbed by the environment transforming it to $\xi(\vec{f}_{T},t)$. The sensor will maximally absorb the photon when its control field $V(\vec{f},t)$ matches the incoming pulse. (b) The Raman model inside the actuator and sensor. It includes a three-level atom enclosed in a singled-sided cavity with decay rate $\kappa$. A strong, but highly detuned classical driving field couples the two long-lived ground states $\ket{g}$ and $\ket{e}$ via the virtual transition with the radiative state $\ket{b}$. Single photons are then emitted into the output mode $a_{o}(t)$ with a temporal profile determined by the control. In the detector model this process is reversed and an incoming photon is perfectly absorbed when a single photon has the same temporal shape as the control field.}
    \label{fig:fig1}
\end{figure*}

In this article, we consider a simple optical agent which utilizes the temporal profile of light to probe a very restricted environment composed of a single optical element such as an optical cavity.
This optical element unitarily transforms the temporal profile of the probe. The agent's objective is to learn the effect of this transformation, and thus the environmental parameters \footnote{We could further consider multiple optical elements and more complex combinations of optical systems.}
In our conception, the quantum part of our agent is due to the possibility of quantum actuators and quantum sensors. The actuators provide single photon sources to probe the world; the sensors capture information about the returning optical pulse through its temporal profile. The processing steps are entirely classical, putting our agent in the quantum-classical (QC) category of quantum agents \cite{Dunjko_Quantum_2016}. We compare this QC agent to a wholly classical agent that uses weak coherent states---with mean photon number of 1.
An example of our conception is depicted in Fig. \ref{fig:fig1}(a). 

Single photons are highly non-classical states of light, unlike coherent light pulses \cite{walls_quantum_2008} which can be attenuated to the single photon limit, containing on average one photon but with Poisonnian intensity fluctuations \cite{walmsley_quantum_2015, milburn_quantum_2015}. 
The agent's sensor---in both the classical and quantum model---is comprised of a single photon detector which has two states, accept (click) or error (no-click).
There are a number of demonstrated schemes for single photon sources and detectors \cite{nisbet-jones_highly_2011, hadfield_single-photon_2009, averchenko_temporal_2017}. 
Here we employ a three-wave-mixing Raman transition for each \cite{james_atomic-vapor-based_2002, kuhn_deterministic_2002}. 

Lastly, the agent learns by updating its prediction and minimizing the probability of detecting an error.
The measured overlap between the agent’s prediction and observations defines a cost function $C$, with the learning rate $L$ that is fixed by the physical parameters of the system.
A schematic of our model is depicted in Fig.\,\ref{fig:fig1}(a).
The detailed description of the agent's sources and detectors presented here allows us to study the thermodynamic behavior of our agent as it learns. We show that the agent's learning maximizes the work done by the incoming pulse on the sensor and, subsequently the free energy available for learning. 

The rest of the article is organized as follows: In the first section we introduce the three-wave-mixing Raman model, using it to realize the actuators and sensors in our quantum agent. We then provide a brief description of the classical agent, again using the Raman model. 
This section is followed by a discussion of the agent's learning abilities. In the final section we analyze the thermodynamic behavior of the agent using a very simple example: the agent must learn an exponentially decaying emission profile, parameterised by a linewidth and detuning of an unknown cavity with respect to the pulse carrier frequency. Regardless of this simple example, our analysis holds for much more complex unitaries and optical pulse shapes.

\section{Raman Model}

In the Raman model, a three-level atom is placed inside a single-sided cavity.
Two long-lived states $\ket{g}$ and $\ket{e}$ are coupled by a third radiative state $\ket{b}$.
A strong classical electromagnetic control pulse $E(t)$ is applied to the ground state $\ket{g}$ at frequency $\Omega$.
The pulse is detuned from the atomic transition $\ket{g}\rightarrow\ket{b}$, ensuring that the radiative state is never significantly populated.
The resonance condition is $\Omega - \omega_{a} = \omega_{\sigma}$, where $\hbar \omega_{\sigma}$ is the energy difference between $\ket{g}$ and $\ket{e}$.
The transition $\ket{g}\rightarrow\ket{e}$ is mediated by the emission of a photon at frequency $\omega_{a}$ as depicted in Fig.\,\ref{fig:fig1}(b).

In the interaction picture under the rotating wave approximation, the Hamiltonian describing this interaction is
\begin{equation}
H_{\mathrm{a}} = \hbar E(t) a^{\dagger} \sigma_{+} + \hbar E^{*}(t)\sigma_{-}a\,,
\end{equation}
where $a^{\dagger}$ and $a$ are the internal cavity mode creation and annihilation operators, and $\sigma_{+}$ and $\sigma_{-}$ are the raising and lowering operators in the subspace formed $\ket{g}$ and $\ket{e}$.
At a finite temperature, the evolution of the cavity-atomic joint state is governed by the master equation (ME)
\begin{align}
\label{eq:Master}
    \frac{d\rho}{dt} = - i \left[H_{\mathrm{a}}, \rho\right] + 
    \kappa\left(\nbar + 1\right) \mathcal{D}[a]\rho + \kappa \nbar \mathcal{D}[a^{\dagger}]\rho\,,
\end{align}
where $\kappa$ is the decay rate of the cavity mode into the environment, $\nbar=(e^{\mu_{a}}-1)^{-1}$ is the mean photon number in the environment with Boltzmann factors $\mu_{i} = \hbar \omega_{i}/k_{B} T$, and $\mathcal{D}[a]\rho = a \rho a^{\dagger} - \{a^{\dagger} a, \rho\}/2$ is the Lindblad dissipator.
We have neglected decay between $\ket{e}\rightarrow\ket{g}$ with the assumption that the atomic decay rate is much slower than the intra-cavity mode $\kappa$. 
We further assume the atom-cavity system is initially in thermal equilibrium with the environment, and thus it is in a separable Gibbs state, $\rho_{\mathrm{sys}} = \bar{\rho}_{a} \otimes \bar{\rho}_{\sigma}$.
Thermal equilibrium ensures $\mu_{a} = \mu_{\sigma}$.

\subsection{Quantum Actuator (Single Photon Source)}

A reliable quantum actuator requires two things; sufficient control over the shape of the output field $a_{o}(t)$ and the photon number $a_{o}^{\dagger}(t)a_{o}(t)$.
Such requirements ensure the agents ability to control both the intensity and optical coherence of its probe precisely and repeatedly.
To compute the output of these quantities, we use the quantum Langevin equations to describe the stochastic evolution of the intracavity $a(t)$ in terms of the input $a_{i}(t)$ and output fields $a_{0}(t)$ \cite{gardiner_input_1985, walls_quantum_2008}
\begin{align}
\label{eq:intracavity}
    &\frac{d a}{dt} = - i E(t)  \sigma_{+}  - \frac{\kappa}{2} a   + \sqrt{\kappa} a_{i} \,.
\end{align}
Solving the Langevin equation given the initial condition that $\langle a(0) \rangle = \langle a_{i}(0) \rangle = 0$ and making use of the input-output relation $a_{i}(t) + a_{o}(t) = \sqrt{\kappa}a(t)$, we find that the output field is---on average---the convolution of the cavity response and the product of the control field amplitude and atomic polarization 
\begin{equation}
    \langle a_{0}(t) \rangle = - i\sqrt{\kappa} \int_{0}^{t}dt' \exp(\kappa (t'-t)/2)E(t')\langle \sigma_{+}(t')\rangle\,.
\end{equation}
Therefore, controlling the shape of the classical drive $E(t)$ shapes the overall output of the agent's actuator.

Likewise the photon flux emitted from the agent's actuator and into the output mode can be ascertained using the input-output relations
\begin{equation}
\label{eq:output_r}
    \langle a_{o}^{\dagger} a_{o} \rangle = \kappa \langle a^{\dagger} a \rangle -\sqrt{\kappa}\langle a^{\dagger}a_{i} + a_{i}^{\dagger}a \rangle + \langle a_{i}^{\dagger}a_{i}\rangle  \,,
\end{equation}
where $\langle a_{i}^{\dagger} a_{i} \rangle = \nbar$ is the mean photon number of the environment. A reliable single photon source will operate in the limit of a large cavity decay rate $\kappa$ and thus preferentially emit into the environment rather than be coherently absorbed by the atom.
In this limit, we can adiabatically eliminate the cavity dynamics such that the cavity dynamics remain largely constant, which from Eq. (\ref{eq:intracavity}) yields
\begin{equation}
\label{eq:adiabatic}
    a \rightarrow -\frac{2iE(t)}{\kappa}\sigma_{+} + \frac{2 }{\sqrt{\kappa}}a_{i}\,.
\end{equation}
After adiabatically eliminating the cavity dynamics, the output photon flux is given by 
\begin{equation}
\label{eq:output}
    \langle a_{o}^{\dagger}(t) a_{o}(t) \rangle =  \nbar+ \frac{2i}{\sqrt{\kappa}}\langle E(t) a_{i}^{\dagger} \sigma_{+} - E^{*}(t) \sigma_{-} a_{i}\rangle \,.
\end{equation}
To determine the mean photon number in the output mode, we must therefore find an expression for the second term.

To find this term, we start by rewriting the master equation (\ref{eq:Master}) for the atomic system by replacing $a$ in the adiabatic limit
\begin{equation}
\label{eq:atomic}
  \mathcal{L}\rho^{(\sigma)} = \frac{4(\nbar + 1)I(t)}{\kappa}\mathcal{D}[\sigma_{+}]\rho^{(\sigma)} + \frac{4 \nbar I(t)}{\kappa}\mathcal{D}[\sigma_{-}]\rho^{(\sigma)}\,,
\end{equation}
where $\rho^{(\sigma)}$ describes the quantum state of the atom alone and $I(t) = \vert E(t) \vert^{2}$.
The dynamics of the atom appear to spontaneously absorb and emit photons according to the mean-photon number of the environment $\nbar$ and the intensity of the control pulse $I(t)$.

We derive the quantum Langevin equation for $\sigma_{+}$ using the input mode $a_{i}$, which yields the general solution to $\sigma_{+}(t)$ 
\begin{equation}
\label{eq:sigmaplus}
    \sigma_{+}(t) = - \frac{2 i}{\sqrt{\kappa}} \int_{0}^{t} dt' e^{\frac{\kappa}{2}\left(t - t'\right)}\sigma_{z}(t')a_{i}(t') E^{*}(t')
\end{equation}
Multiplying this expression by $a_{i}^{\dagger}$ from the right and making use of the commutation relation $[a_{i}(t),a_{i}^{\dagger}(t')] {=} \delta(t-t')$ and the integral identity $\int_{0}^{t}\delta(t-t')g(t')dt' = g(t)/2$, we find the general solution 
\begin{equation}
    \frac{2i}{\sqrt{\kappa}}\langle E(t) a_{i}^{\dagger} \sigma_{+} - E^{*}(t) \sigma_{-} a_{i}\rangle =  \frac{4 I(t)}{\kappa} \langle \sigma_{z}(t)\rangle \,.
\end{equation}
Rewriting Eq. (\ref{eq:output}), we obtain the expression for the mean photon number in the output mode
\begin{equation}
    \langle a_{o}^{\dagger}(t) a_{o}(t) \rangle = \nbar +  \frac{4 I(t)}{\kappa}\langle \sigma_{z}(t)\rangle\,.
\end{equation}
Finally we can find the behavior of $\langle \sigma_{z}(t) \rangle$.
Making use of the fact that $\sigma_{z} = 1 - 2 P_{g}(t)$, we can find the evolution of $\sigma_{z}$ by computing the probability of measuring a photon in the ground state $P_{g}(t)$.
Thus, in the Heisenberg picture $P_{g}(t)$ evolves according to Eq. (\ref{eq:atomic}), resulting in
\begin{equation}
\label{eq:ground}
\frac{dP_{g}(t)}{dt} =  -\frac{4I(t)}{\kappa}(2\nbar + 1 )P_{g}(t)  + \frac{4I(t)\nbar }{\kappa}  
\end{equation}
which has the following general solution---given the initial condition $P_{g}(0) {=} (1+e^{-\mu_{\sigma}})^{-1}$:
\begin{equation}
\label{eq:ground_1}
    P_{g}(t) = \frac{e^{-\tau}(1+\nbar) + \nbar}{2\nbar + 1}-\frac{e^{-\tau}}{1 + e^{\mu_{\sigma}}}\,.
\end{equation}
Here $\tau = \left(4/(2\nbar + 1)\kappa\right) \int_{0}^{t}dt' I(t')$ and approaches $0$ in the long-time limit at zero temperature, corresponding to a perfect emission.
In the long-time limit where $t\gg 0$, the polarization of the atom $\langle \sigma_{z}(t)\rangle$ becomes constant
\begin{equation}
\label{eq:sz}
    \langle \sigma_{z}(t) \rangle = \frac{1}{2 \nbar + 1} = \tanh\left( \frac{\hbar \omega}{2 k_{B} T}\right)\,,
\end{equation}
which describes the mean atomic polarization of a two-level atom in a thermal bath as expected.
The mean photon number that is emitted by the quantum agent's actuator in the long-time limit is given by 
\begin{equation}
    \langle a_{o}^{\dagger}(t) a_{o}(t) \rangle = \nbar +  \frac{4 I(t)}{\kappa}\tanh\left( \frac{\hbar \omega}{2 k_{B} T}\right)\,.
\end{equation}
In conclusion, the Raman model that we considered guarantees the agent's ability to control the output mode precisely and repeatedly, given it has sufficient control of the classical drive $E(t)$.

\subsection{Quantum Sensor (Single Photon Detector)}
We will further assume the actuator pulse returns to the detector after its temporal profile is perturbed unitarily.
{As we detailed in the previous section, the output mode of the single-photon source has a temporal mode defined by the control field $E(t)$. After this pulse interacts with the world, its temporal profile transforms unitarily, $E(t) \rightarrow \xi(\vec{f}_{T}, t)$---which we abbreviate to $\xi(\vec{f}_{T}, t) \equiv \xi$ [Fig.\,\ref{fig:fig1}(a). }
Here the true discoverable parameters $\vec{f}_{T}$ determine the measurable effect of the environment on the pulse.
{If the agent can estimate $\vec{f}_{T}$, then it has `learned' the environment.}
The actuator model can be suitably adapted into a detection model by reversing the process as depicted in Fig.\,\ref{fig:fig1}(a).
The virtual transition between the $\ket{g}$ and $\ket{e}$ is again mediated by a classical control field---denoted $V(\vec{f}, t)$, where $\vec{f}$ are the agent's control parameters determining its control pulse.
We will use the shorthand notation $V(\vec{f}, t) \equiv V$.
In the interaction picture, the Hamiltonian describing the sensor is
\begin{equation}
H_{\mathrm{s}} = \hbar V(\vec{f}, t) a^{\dagger} \sigma_{+} + \hbar V^{*}(\vec{f}, t) \sigma_{-}a\,.
\end{equation}
The sensor requires a source of energy to ensure a population inversion; the excited state $\ket{e}$ is now preferentially populated, creating a negative temperature equilibrium state satisfying $\mu_{a} = - \mu_{\sigma}$ \cite{ramsey_thermodynamics_1956}.
A successful detection occurs when an incoming photon at frequency $\omega_{a}$ is absorbed and deexcites the atom into the ground state $\ket{g}$.
{By tuning the parameters $\vec{f}$, the agent can maximize the probability that the incoming photon is absorbed.}
The state of the atom can then be read out accurately with negligible dissipation using fluorescent imaging \cite{james_atomic-vapor-based_2002}.

We cannot describe the absorption process via the standard master equation Eq. (\ref{eq:Master}), but rather via the Fock state master equations \cite{baragiola_n-photon_2012}.
In this framework, the entire system is described by a joint system $\rho_{\mathrm{joint}} = \rho\otimes \ket{1_{\xi}}\bra{1_{\xi}}$, where the incoming photon is in a single-photon Fock state,
\begin{equation}
    \ket{1_{\xi}} = \int dt \xi(\vec{f}_{T}, t)b^{\dagger}(t)\ket{0}\,,
\end{equation}
where $b^{\dagger}(t)$ is the creation operator of the incoming Fock mode.
For a single-sided cavity, this yields the upwardly coupled master equations
\begin{align}
\label{eq:Fock}
    \frac{d \rho_{m,n}}{dt} = \mathcal{L} \rho_{m,n} + \sqrt{m}\sqrt{\kappa\eta}\xi \left[\rho_{m-1,n}, a^{\dagger}\right]
    \nonumber\\
    +\sqrt{n}\sqrt{\kappa \eta}\xi^{*}\left[ a, \rho_{m,n-1}\right]\,,
\end{align}
where $\eta$ is the quantum efficiency of the detector and $m$ and $n$ are integers.
For a single photon Fock state, $\ket{1_{\xi}}$, $m,n\in \{0,1\}$.
The equation for ${\rho}_{0,0}$ is identical to the vacuum master equation ({\ref{eq:Master}}) and can be solved in principle.
The diagonal elements ${\rho}_{n,n}$ are initialized with ${\rho}_{\mathrm{sys}}(0)$ whereas the off-diagonal elements are initialized to zero.
Only the top density operator ${\rho}_{11}$ is required to compute expectation values. 

In the single-photon Raman model, a successful detection occurs when the atom is measured in the ground state $\ket{g}$.
For example, this could be done accurately with negligible dissipation using fluorescent imaging \cite{james_atomic-vapor-based_2002}.
We can now repeat our analysis in the previous section assuming it is in thermal equilibrium with a thermal bath of mean photon number $\nbar$.
The superoperator further includes thermal excitations, yielding the master equation
\begin{align}
\label{eq:met}
    \frac{d \rho}{dt} = - i \left[H_{s}, \rho\right]+\kappa(\nbar+1) \mathcal{D}[a]{\rho} + \kappa \nbar \mathcal{D}[{a}^{\dagger}]{\rho}\,,
\end{align}
We make the same assumptions, as before including the adiabatic approximation, similar to Eq. (\ref{eq:adiabatic}).
We compute the quantum Langevin equation for the incoming Fock-mode resulting in 
\begin{equation}
    a = -\frac{2iV\sigma_{+}}{\kappa} - \frac{2 \xi}{\sqrt{\kappa}}\,.
\end{equation}
After substituting this back into Eq. (\ref{eq:met}), we eliminate the cavity dynamics entirely yielding the atomic master equation for the detector,
\begin{equation}
\label{eq:met-atom}
    \mathcal{L}\rho^{(\sigma)}= \frac{4(\nbar + 1)I_{V}}{\kappa}\mathcal{D}[{\sigma}_{+}]{\rho}^{(\sigma)} + \frac{4 \nbar I_{V}}{\kappa}\mathcal{D}[{\sigma}_{-}]{\rho}^{(\sigma)}\,,
\end{equation}
where $I_{V} = \vert V\vert^{2}$.
This master equation now corresponds to the vacuum---the lowest $\rho_{0,0}$---term in our coupled master equations. 
Thus, we recast Eq. (\ref{eq:Fock}) obtaining the Fock state atomic master equation 
\begin{align}
\label{eq:fock_me}
    \frac{d{{\rho}}^{(\sigma)}_{n,m}}{dt} = \mathcal{L}\rho^{(\sigma)}_{n,m}  \frac{2i\sqrt{\eta}}{\sqrt{\kappa}}
    &\left(\sqrt{n}\xi V^{*}\left[{\rho}^{(\sigma)}_{n-1,m}, {\sigma}_{-}\right]\right. \nonumber
    \\
    &\left.- \sqrt{m}\xi^{*} V \left[{\sigma}_{+}, {\rho}^{(\sigma)}_{n,m-1}\right] \right)\,.
\end{align}
A successful detection will occur when the incoming Fock state excites the atom into the ground state. Thus, we seek to maximize $P_{g}(t)$ which evolves according to
\begin{align}
\label{eq:PG_fock}
     \frac{dP_{g}(t)}{dt} &=  -\frac{4I_{V}}{\kappa}(2\nbar + 1 )P_{g}(t)  + \frac{4I_{V}\nbar }{\kappa}\nonumber \\
     &+ \frac{2i \sqrt{\eta} V^{*} \xi}{\sqrt{\kappa}} 
    \langle {\sigma}_{-}\rangle_{01} - \frac{2i \sqrt{\eta} V \xi^{*}}{\sqrt{\kappa}}\langle {\sigma}_{+}\rangle_{10}\,.
\end{align}
We must now find $\langle {\sigma}_{-}\rangle_{01}$ which can also be computed using the Eq. (\ref{eq:fock_me})
\begin{equation}
        \frac{d \langle {\sigma}_{-}\rangle_{01}}{dt} = -\frac{2(2\nbar+1)I_{V}}{\kappa}\langle {\sigma}_{-}\rangle_{01}- \frac{2i\xi^{*}V}{\sqrt{\kappa}}\langle {\sigma}_{z}\rangle_{00}\,.
\end{equation}
Lastly, we must find the evolution of $\langle {\sigma}_{z}\rangle_{00}$.
Given that the sensor is maintained in a negative temperature state, the probability of finding it in the ground state when the environment is time independent for an incoming vacuum, is $P_{g}(t)_{00} = (1 + e^{\mu_{\sigma}})^{-1}$.
Using the initial condition $\langle {\sigma}_{-}(0)\rangle_{01} = 0$, we obtain the general solution
\begin{equation}
    \langle {\sigma}_{-}(t)\rangle_{01} {=} - \frac{2i}{2\nbar +1}\sqrt{\frac{\eta}{\kappa}}\int_{0}^{t} e^{(\tau'-\tau)}V(\vec{f}, t')\xi^{*}(\vec{f}_{T},t') dt'\,,
\end{equation}
where $\tau = 2\int_{0}^{t}dt' (1+2n) I_{V}(t')/k$. 
Substituting this result into our differential equation for $P_{g}(t)$, i.e, Eq. (\ref{eq:PG_fock}), we can compute the general solution to this differential equation using the initial condition of the atom,
\begin{equation}
    P_{g}(t) {=}\frac{1}{2\nbar + 1} \left(\nbar + \frac{4\eta }{\kappa}\left| \int_{0}^{t} dt' e^{(\tau'-\tau)} V(\vec{f}, t')\xi^{*}(\vec{f}_{T},t')\right|^{2}\right)\,.
\end{equation}
If we assume that $\kappa$ is large and the response function becomes close to instantaneous, then we can simplify this expression to
\begin{equation}
\label{eq:probQ}
    P_{g}^{(Q)}(t) = \frac{\nbar}{1 + 2\nbar} + \frac{4 \eta \Gamma}{\kappa} \tanh \left(\frac{\mu_{\sigma}}{2}\right)\,,
\end{equation}
where  $\Gamma = \left|\int dt' V^{*}(\vec{f}, t')\xi(\vec{f}_{T}, t') \right|^{2}$ and we have included the superscript $(Q)$ to indicate the quantum agent. 
Thus, the probability of measuring the atom in the ground state depends on the thermal state of the atom, and maximising the overlap between $V$ and $\xi$.
Moreover, the first and second terms correspond to the conditional probability of the atom transitioning to the ground state due to absorption of a thermal photon and incoming Fock state, respectively.

\begin{figure}
    \centering
    \includegraphics[width=\columnwidth]{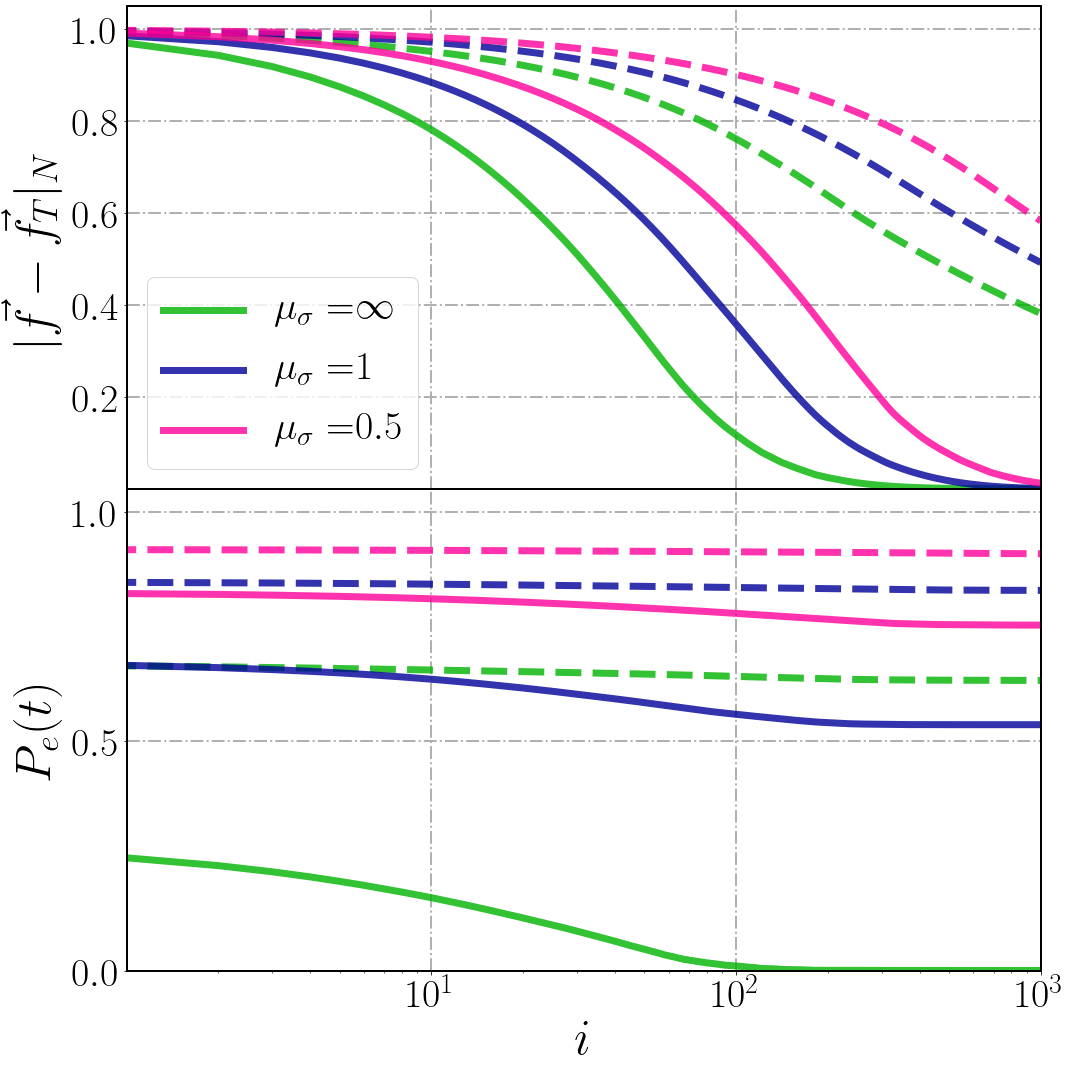}
    \caption{Top: The normalized parameter difference---each parameter is dimensionless between $0$ and $1$---of the agent's prediction and observation, $|\vec{f}-\vec{f}_{T}|_{N}$, defined as a function of iterations $i$. The quantum agent (solid lines) outperforms the classical agent (dashed lines) at all temperatures, excluding the infinite temperature limit $\mu_{\sigma} \rightarrow 0$. Bottom: The probability of measuring an error in the incoming pulse $\xi(\vec{f}_{T},t)$. When the estimate is incorrect, $\Gamma <1$, the probability of the atom deexciting into the ground state is not guaranteed. As the agent's estimate improves, the probability of obtaining an error decreases.}
    \label{fig:learning}
\end{figure}

\subsection{The Classical Agent}

In the classical model, the agent realizes it's actuator by pumping directly into the environment with a coherent pulse with a mean photon number of $1$.
We choose the weak coherent state as the optimal classical actuator because it overlaps minimally with the vacuum at $\langle n\rangle =1$. Other Gaussian states such as displaced squeezed states, or thermal states, exhibit larger overlap with the vacuum at $\langle n \rangle =1$ with a larger variance in the photon number \cite{walls_quantum_2008}. As we will see, these properties hinder the performance of the classical agent by increasing the error rate of the detector. Thus, a coherent state is the optimal choice of probe for a classical agent.

For a coherent state, the mean output mode in this model is $\langle a_{0}(t)\rangle {=} -i\int_{0}^{t}dt'E(t')$ with a mean photon number $\langle a_{0}^{\dagger}(t)a_{0}(t)\rangle {=} \nbar + \int_{0}^{t}dt'\vert E(t')\vert^{2}$.
For the agent's classical detector, it uses the same single-photon detector as the quantum model, but replaces the incoming single-photon pulse with a coherent state with temporal shape $\xi(\vec{f}_{T},t)$. 
Replacing the incoming Fock state in the Fock state master equation (\ref{eq:Fock}) with a coherent state leads to a new master equation equivalent to Eq. (\ref{eq:met}) but with an additional coherent drive term proportional to $\xi(\vec{f}_{T},t)$,
\begin{align}
\label{eq:Master_drive}
    \frac{d\rho}{dt} = - i \left[H_{\mathrm{s}}, \rho\right] + 
    \kappa\left(\nbar + 1\right) \mathcal{D}[a]\rho + \kappa \nbar \mathcal{D}[a^{\dagger}]\rho \nonumber \\ + i\sqrt{\eta \kappa}\left[\xi^{*}a - \xi a^{\dagger}, \rho \right]\,.
\end{align}
By repeating the analysis as the quantum detector, we obtain---in the long-time and large $\kappa$ limit---the conditional probability of measuring the atom in the ground state from a coherent pulse as
\begin{align}
\label{eq:probC}
    P_{g}^{(C)}(t) = \frac{\nbar}{2\nbar + 1} + \frac{4 \eta \Gamma}{\kappa}  e^{-4\eta \Gamma/\kappa}\tanh \left(\frac{\mu_{\sigma}}{2}\right)\,,
\end{align}
where we have included the $(C)$ superscript to indicate that this is the classical agent.
Thus, the primary difference between the classical model and the quantum model is due to the exponential dependence on overlap $\Gamma$, which is due to the intensity fluctuations in the coherent field \cite{walls_quantum_2008}. 

\section{Learning}
With the hardware of our agent specified, we can move onto describing the agent's software and its ability to learn.
We assume that the agent and the environment are materially identical---otherwise known as the principle of requisite variety \cite{ashby_introduction_1960}---so matching $V(\vec{f},t)$ and $\xi(\vec{f}_{T},t)$ corresponds to matching the parameters $\vec{f}$ and $\vec{f}_{T}$ 
\footnote{Note, this \emph{a priori} assumption endows the agent with some initial information about its unknown environment. In principle, the pulse shape $V(\vec{f}, t)$ could be generated by a device that can universally represent all pulse shapes, and the task of learning would be to simply match the pulse and maximize the likelihood of a detection event. }.
Here, we will describe how the agent can realize a simple form of gradient descent using the measured error rate from its single photon detector.

{Our agent can only use the state of its detector to infer information about the environment. For each detection event $j$, the agent measures the detector in the ground or excited state, both of which register a classical bit of information $x_{j}\in \{0,1\}$.
When the agent's control pulse $V(\vec{f},t)$ perfectly overlaps the world's $\xi(\vec{f}_{T},t)$, the probability of measuring an error (excited state) $P_{e}(t) = 1-P_{g}(t)$ is minimized. 
Given the probabilistic nature of quantum measurement, $x_{j}$ is a binary random variable.
The agent stores $N$ samples of $x_{j}$ in memory, creating a bit string which it uses to estimate the \emph{mean error rate} $\bar{x}_{N} = \sum_{j=0}^{N} x_{j}/N$---which in the limit of large $N$ approaches $\bar{x}_{N} \rightarrow P_{e}(t)$. We will define a single experimental iteration $i$ as the collection of $N$ measurement events. Thus, for each experimental iteration $i$, the agent estimates a mean error rate $\bar{x}_{N}^{(i)}$. }

{For the agent to learn, it must minimize $\bar{x}_{N}^{(i)}$ and, subsequently maximize $\Gamma$. 
Between experimental runs, the parameters $\vec{f}$ are updated.
Using the chain rule, we can define the rate of change in the mean error rate, for an incremental change $di$, as
\begin{equation}
    \frac{d \bar{x}_{N}^{(i)}}{d i} = \frac{d\vec{f}}{d i}\cdot \vec{\nabla}_{\vec{f}} P_{e}(t)\,.
\end{equation}
Consequently, the mean error rate will be minimized when either the parameters no longer update $d\vec{f}/di = 0$, or the probability of measuring an error $P_{e}(t)$ has been minimized. 
This provides an intuitive mathematical definition of learning: the agent learns by minimizing the number of measured errors.}

{There are many algorithms that are capable of updating the parameters $\vec{f}$.
We choose to use gradient descent (GD) as a simple example which updates $\vec{f}$ numerically 
\begin{equation}
     \vec{f}_{i+1} = \vec{f}_{i} + L \vec{\nabla}_{\vec{f}} P_{e}(t)\,,
\end{equation} 
where $L$ is the learning rate.
The learning rate will be constrained by the physical parameters of the detector, i.e, by its thermal state $\mu_{\sigma}$ and quantum efficiency $\eta$ as specified in the probability distributions given by Eq. (\ref{eq:probQ}) and Eq. (\ref{eq:probC}). 
Also, each experimental run is repeated every $s$ seconds, and thus the time between each control integration is $Ns$. 
This further bounds the learning rate below $1/Ns$.}

As an example, suppose the control $V(\vec{f},t)$ and input pulses $\xi(\vec{f}_{T},t)$ are exponentially decaying temporal modes of the form $\sqrt{\gamma}\exp \left(-\gamma t/2 + i \Delta t\right)$ for $t\geq 0$ generated by an optical cavity with linewidth $\gamma$ and detuning $\Delta$.
{Here the agent must determine the linewidth and detuning of the world's cavity $\vec{f}_{T} = (\gamma_{T}, \Delta_{T})$.}
The convergence between the true parameters $\vec{f}_{T}$ and prediction $\vec{f}$ can be monitored via the normalized Euclidean distance $\vert \vec{f}-\vec{f}_{T}\vert_{N}$ shown in Fig. \ref{fig:learning}.
Assuming both models have a fixed learning rate $L$, the quantum agent (solid lines) outperforms the classical agent (dashed lines) at all temperatures, converging on the true estimates an order of magnitude faster.
Thus, in this scenario, the quantum agent outperforms the classical agent.

\begin{figure}
    \centering
    \includegraphics[width=\columnwidth]{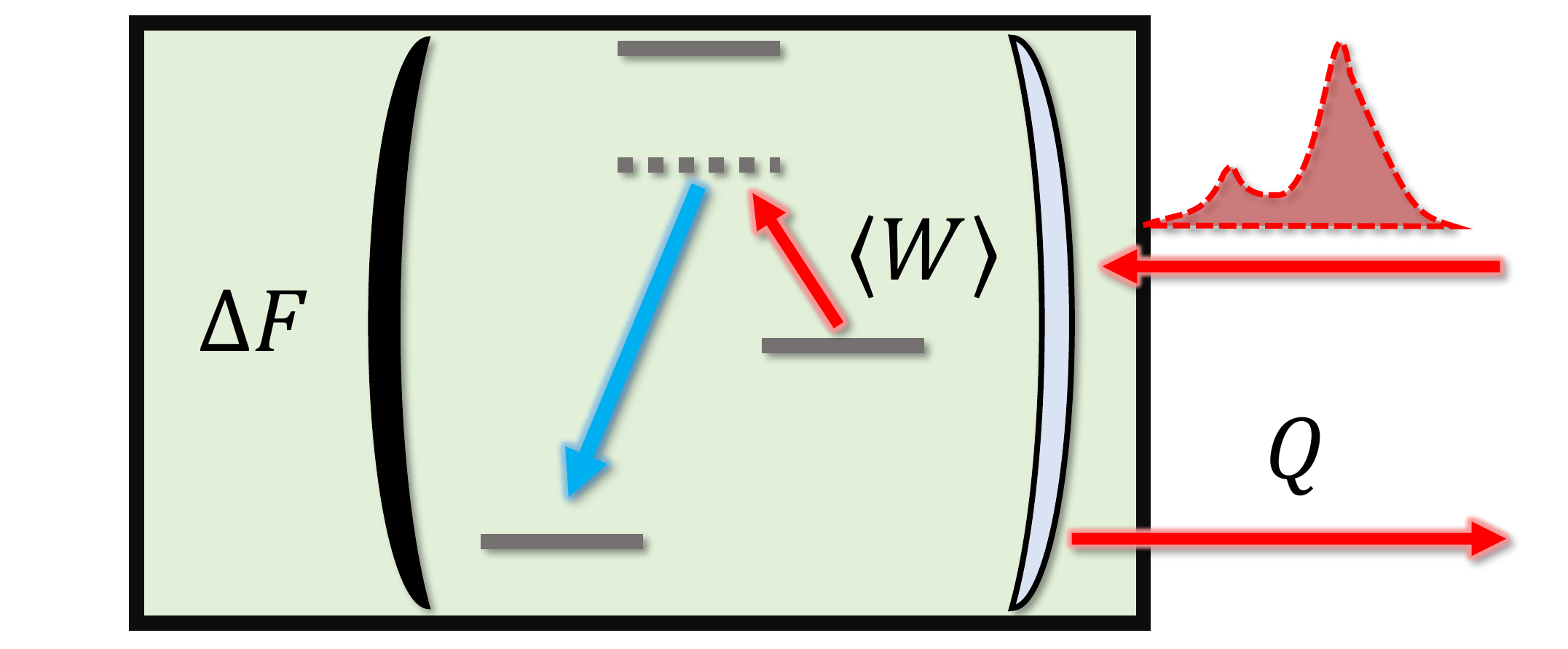}
    \caption{The free energy $\Delta F$ of the detector increases if the photon does work $\langle W \rangle$ on the atom by stimulating a transition. If the photon is not absorbed, it is reflected back into the environment as heat $Q $.}
    \label{fig:thermal_optics}
\end{figure}

\section{Thermodynamics}

{As we have discussed for our model, learning requires emulating the environment using sensory data. 
Learning requires the agent consumes energy; it requires a power source.
Here, we assume our agent has access to a large reservoir of free energy which it uses to probe the environment via single photons or coherent states. 
It converts this free energy into light, which subsequently does work on the detector, updating the state of the detector and enabling the agent to learn.
We can study the thermodynamics of our agent through the lens of quantum machines and molecular systems \cite{quan_quantum_2007, seifert_stochastic_2012}.
A seminal result in non-equilibrium thermodynamics is the Jarzynski equality \cite{jarzynski_equilibrium_1997} 
\begin{equation}
\label{eq:jarzynski}
    \langle e^{-W \beta} \rangle = e^{-\Delta F \beta}\,,
\end{equation}
which relates the free energy difference $\Delta F$ between two thermodynamic states to the irreversible work $W$ required to drive it between the two at inverse temperature $\beta = 1/k_{B}T$. }

{Given the probabilistic nature of our agent, we can compute the average work done $\langle W \rangle/\mu_{a}$ when a signal photon with energy $E=\hbar \omega_{a}$ is absorbed by the detector.
Absorbing the incoming photon updates the state of the detector and subsequently provides the agent with useful information that can be used for learning. 
Moreover, if the photon is reflected by the detector, it is unchanged and dissipated as heat $ Q  = \langle W \rangle - \Delta F$ \cite{seifert_stochastic_2012}.
We can further compute the change in free energy $ \Delta F $ of the detector via the Jarzynski equality Eq. (\ref{eq:jarzynski}) of the detector after the photon is absorbed or reflected.
Thus absorbing a photon increases the free energy of the detector as we would expect.}

\begin{figure}
    \centering
    \includegraphics[width=\columnwidth]{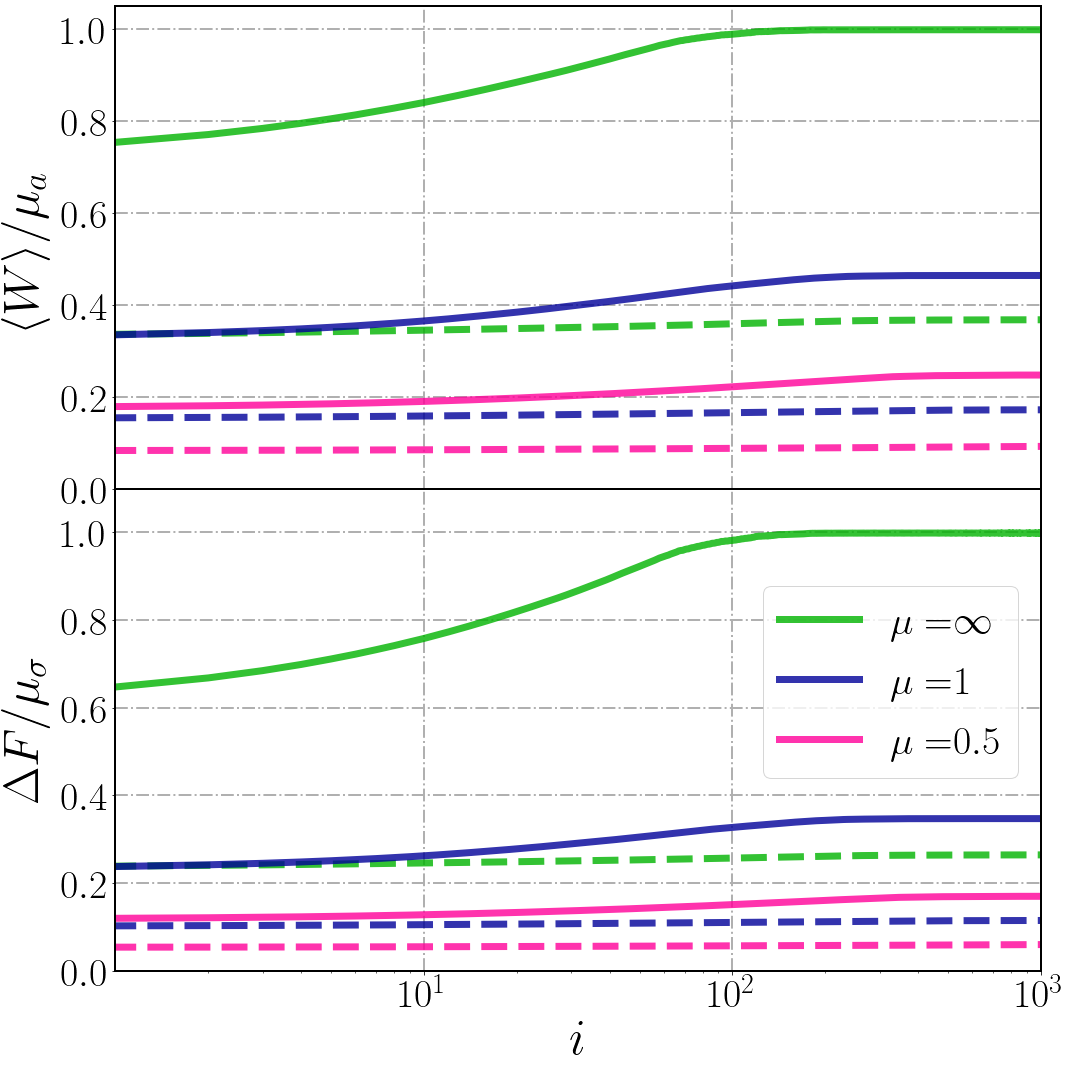}
    \caption{ Top: The scaled average work done by the incoming mode on the atom $
    \langle W \rangle /\mu_{\sigma}$ and the change of free energy in the atom $\Delta F/\mu_{\sigma}$. The energy is transferred between actuator and sensor after interacting with the environment. As the agent's estimate $\vec{f}$ improves, the change in free energy in the detector and the useful work that is done are maximized. As the temperature increases the amount of energy exchanged decreases in both models.}
    \label{fig:work}
\end{figure}

{In the example we considered previously, we find that $\langle W\rangle $ and $\Delta F$ continue to increase as $\vec{f}$ approaches $\vec{f}_{T}$, depicted in Fig. (\ref{fig:work}).
At zero temperature, $\mu_{a} = \infty$ and $\eta=1$, we have $\Delta F \rightarrow \langle W \rangle$ as the agent's estimate of the world approaches the true value $\vec{f}\rightarrow\vec{f}_{T}$.
As the temperature increases, $\Delta F/\mu_{a}$ decreases, indicating less work is reliably used for learning and is being dissipated back into the environment, and hence $\langle W\rangle > \Delta F$.
Thus, our model shows that learning maximizes the work done by the sensors pulse on the detector after it has interacted with the environment. Our result corroborates the recent results in Ref. \cite{boyd_thermodynamic_2020}, which showed that learning maximizes the work production in a Maxwell's demon mode. As such, learning may be conceived as an out-of-equilibrium thermodynamic process, consuming power at each iteration to update its parameters $\vec{f}$ with each experimental iteration.}

Lastly, the classical agent's capacity to convert work $\langle W \rangle$ into learning is hindered by the fact that weak coherent states are primarily dominated by vacuum i.e no photon was emitted from the actuator. 
When the error rate---$P_{e}(t)$---between the classical and quantum models is roughly equivalent, the convergence rate in the quantum agent is roughly an order of magnitude higher than the classical agent, shown in Fig. (\ref{fig:learning}). 
This is due to the variable learning rate, which approaches $0$ as $\Gamma\rightarrow1$ since $\nabla_{\vec{f}}P_{g}^{(C)}(t) \propto (1-4 \eta\Gamma/\kappa)e^{-4 \eta \Gamma/ \kappa}$. 
As the classical agent converges on the true values, the intrinsic uncertainty in the probe's photon number makes it more difficult to resolve the smaller differences between the prediction and observation. 
This limitation is not present in the quantum model, which is limited only in the estimate of the error rate i.e $1/\sqrt{N}$ \cite{giovannetti_advances_2011, degen_quantum_2017}.
To conclude this section, the quantum agent utilizes its resources more effectively than the classical agent. This results in a lower error rate and lower likelihood that the average work done by the actuator on the sensor is dissipated back into the environment.

\section{Conclusion}

{In this article, we have presented a simple toy model of a quantum agent.
The agent uses free energy to probe its environment via optical pulses generated by its actuator: a three-wave mixing Raman process.
After interacting with the environment, the pulse is perturbed and returns to the agent where it was measured by its sensor.
The likelihood that the pulse is absorbed by the detector is maximized when the agent has ``learned'' the effect of the environment on the pulse shape.
Thus, by maximizing the probability of a detection of the environment, the agent can learn its environment.
Furthermore, maximizing this likelihood of a detection increases the work done---and subsequently the free energy---by the pulse on the detector.
When the returning pulse is not absorbed, it is dissipated back into the environment as heat.
Thus, it is conceivable that an agent's capacity to learn its environment improves the overall thermodynamic efficiency of transmitting energy from the actuator to its sensor.
From this perspective, learning may be conceived of as an out-of equilibrium thermodynamic process, requiring power to update the agent's estimate of the world.
While our model is simple, it hints at a more subtle connection between learning and thermodynamics which is worth exploring in further detail.}

\section{acknowledgements}
This work was supported by FQXi FFF Grant No. FQXi-RFP-1814 and the Australian Research Council Centre of Excellence for Engineered Quantum Systems (Project No. CE170100009).

\bibliography{Agents_2020}

\newpage

\end{document}